# On the Dynamical Behaviour of Cellular Automata

by Xu Xu* , Yi Song **& Stephen P. Banks***
Department of Automatic Control and Systems Engineering,
University of Sheffield,
Mappin Street, Sheffield, S1 3JD

**Abstract:** In this paper we study the dynamics of 1- and 2- dimensional cellular automata, using a 2-adic representation of the states, we give a simple graphical technique for finding periodic solutions. We also study the continuity properties of the associated 2-adic system and show how to compute the entropy.

## 1. Introduction

Cellular automata have been studied for many years, largely from the point of view of numerical experiments which seek to find the qualitative structure of the complex patterns generated by different rules (see [Wolfram, 1986, Woolters et al, 1990]). They have also been used [*]to model complexity in a wide variety of situations, namely from biological perspective- for example, Conway's game of life, and modeling ant colonies etc (see [Bossomaier and Green, 2000]. On the other hand, there has not been a great deal of analysis of the nonlinear dynamical system defined by local rules, although there is a large literature on the local discrete representation of partial differential equations and their associated cellular dynamics (see [Chow, 2003]). Finding periodic solutions in highly nonlinear systems is extremely difficult, in general. In this paper we shall describe a simple graphical method for determining fixed points and periodic orbits in 1- and 2- dimensional cellular automata. We shall associate with the state of a cellular automaton (which is a 1- or 2- dimensional binary array) a real number in the interval $[0,1)$, by using its 2- adic representation. This will lead to a map $\tilde{N}:[0,1)\rightarrow[0,1)$ which represents the automaton. Periodic points of the automaton are then fixed points of the maps $\tilde{N}^k=\underbrace{\tilde{N}\circ\tilde{N}\circ\cdots\circ\tilde{N}}_{k}$ for some $k$. These will appear as points on the diagonal of the graph of $\tilde{N}^k$. We shall see that this leads to a very simple method for finding periodic (or almost periodic) points of the game of life (or any other cellular automaton). We shall also study the smoothness properties of $\tilde{N}$. We also discuss the entropy of cellular automata, which is essentially a measure of the complexity of the dynamics.

* Email: cop06xx@sheffield.ac.uk
**Email: yi.song@sheffield.ac.uk
***Email: s.banks@sheffield.ac.uk

## 2. One-Dimensional Cellular Automata, Fixed points and Periodic Solutions

A one-dimensional cellular automaton is a discrete nonlinear dynamical system with states given by strings of binary symbols $b = b_1 b_2 \cdots\cdots \in B$ (the length of the strings can be finite or infinite and $B$ denotes the set of all such strings) and the dynamics defined by local rules. By the latter we mean the following:

the dynamics $N$ : if $b' = N(b)$ , then $b'$ is obtained from $b$ by choosing a map $R : S_p \to \{0,1\}$, called the ( $p$ -bit) rule, where $p$ is an odd integer and $S_p$ is the set of all binary strings of length $p$ (so $\# S_p = 2^p$ ), and then setting

$$b_j' = R\left(b_{j-(p-1)/2} b_{j-1} b_j b_{j+1} \cdots b_{j+(p-1)/2}\right).$$

If $b$ is semi-infinite or is finite we must have appropriate boundary conditions. Thus, if , $b = b_1 b_2 \cdots b_k$ then for a 3-bit rule $b_0$ and $b_{k+1}$ are undefined. We can take them as $0$ or $1$ or choose a periodic boundary condition, so that $b_0 = b_k$ and $b_{k+1} = b_1$ . (Similar considerations apply to $p$ -bit rules.) It is well-known that such maps generate complex, self-similar and fractal structures. For example the rule $R : S_3 \to \{0,1\}$ given by

| | R |
|---|---|
| 000 | 0 |
| 001 | 1 |
| 010 | 1 |
| 011 | 1 |
| 100 | 1 |
| 101 | 1 |
| 110 | 1 |
| 111 | 0 |

(2.1)

generates the Sierpinski triangle from a single non-zero bit binary string (see fig.1).

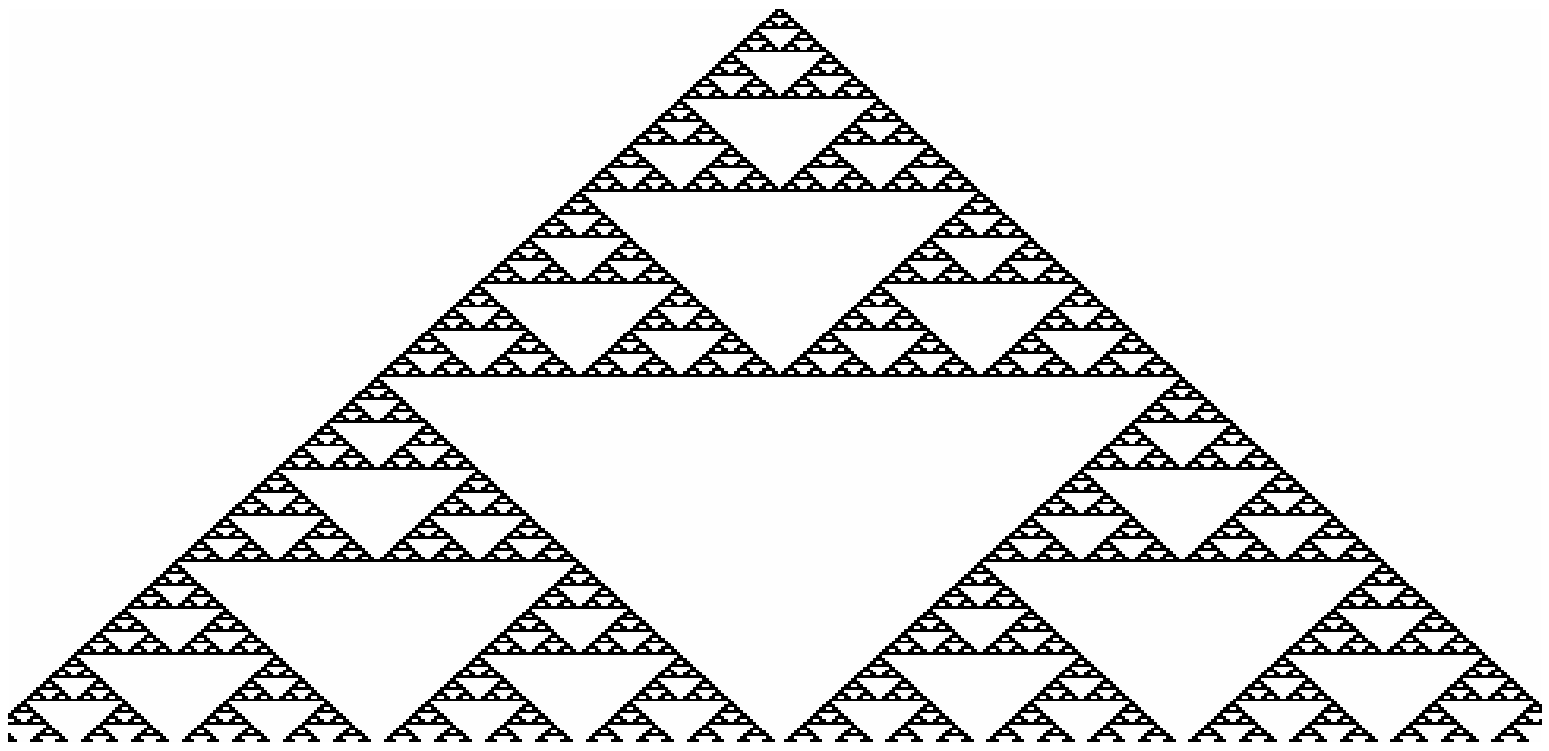

fig.1 Sierpinski Triangle

We shall be interested in the usual dynamical problems (fixed points, periodic orbits, etc) for these systems. First we consider the problem of determining fixed points, i.e. binary strings $b$ such that

$$N(b) = b.$$

To find such points, consider the restriction of the rule $R : S_p \to \{0,1\}$ to the subset $\tilde{S}_p$ of $S_p$ for which

$$R\left(\beta_0 \cdots \beta_{p-1/2} \cdots \beta_{p-1}\right) \to \beta_{p-1/2}$$

(i.e. strings for which $R$ fixes the central bit). Suppose that $K$ of the $2^p$ possible strings $\beta = \left(\beta_0 \cdots \beta_{p-1/2} \cdots \beta_{p-1}\right)$ satisfy this property and label them $v_1, \cdots, v_k$. We shall form an oriented graph $G$ with vertices $V = \{v_1, \cdots, v_k\}$ and oriented edges $\{v_i, v_j\}$ if $v_i$ and $v_j$ 'fit together', i.e. if

$$v_i = \beta_0^i \cdots \beta_{p-1/2}^i \cdots \beta_{p-1}^i$$
$$v_j = \beta_0^j \cdots \beta_{p-1/2}^j \cdots \beta_{p-1}^j$$

then we must have

$$\beta_1^i \cdots \beta_{p-1}^i = \beta_0^j \cdots \beta_{p-2}^j.$$

Now consider a state $b$ which is fixed i.e. $b = N(b)$, so that $b$ is of the form

$$b = b_1 b_2 b_3 b_4 \cdots\cdots$$

where $R(b_{i-1} b_i b_{i+1}) = b_i$.

In the three bit case, for $i = 1, 2, \cdots$, where $b_0 = 0$. We can think of $b$ as a sequence of interlocking 3-bit binary strings:

$$\begin{array}{lllllll} b: & (0 & b_1 & b_2) & & & \\ & & (b_1 & b_2 & b_3) & & \\ & & & (b_2 & b_3 & b_4) & \\ & & & & .. & .. & .. \end{array}$$

(similarly in the p-bit case). Thus, $b$ can be regarded as a path in the graph $G$. Hence the system has a fixed point if and only if there is an infinitely long path in $G$ starting with a vertex of the form $(0 \quad * \quad *)$, where $*$ is either $0$ or $1$. Since $G$ is a finite graph this means that

$G$ must contain cycles. Thus we have

**Lemma 2.1** The one-dimensional automaton defined by the rule $R$ of length $p$ has a fixed point *iff* the graph $G$ associated with it as above has at least one cycle, and the cycle can be reached from a vertex of the form $\left(\underbrace{0,0,\cdots,0}_{p-1/2},*,*,\cdots,*\right)$. The number of distinct fixed points is equal to the number of distinct cycles (which can be $\infty$). □

**Example2.1** For the 'Sierpinski rule' in (2.1) the fixed elements of the rule are

$$
\begin{aligned}
000 &\to 0: \quad v_1 \\
010 &\to 1: \quad v_2 \\
011 &\to 1: \quad v_3 \\
110 &\to 1: \quad v_4
\end{aligned}
$$

Hence, in the case, the graph takes the form

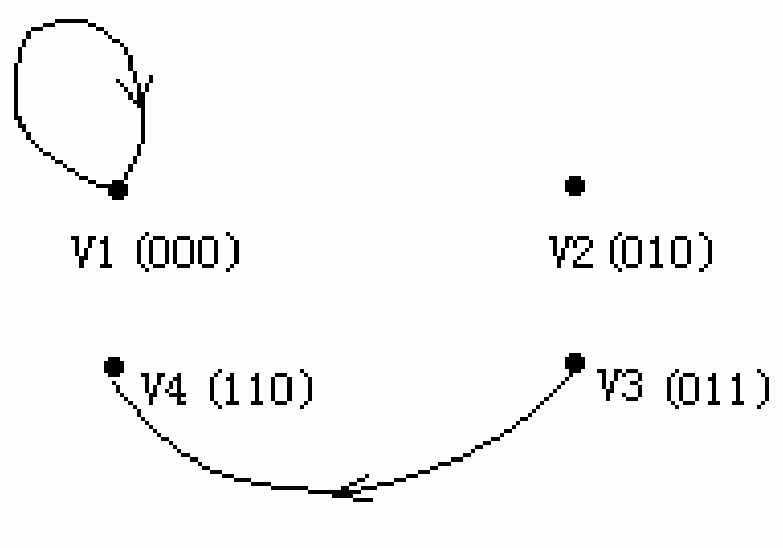

It is clear from the graph that there is a unique fixed point for this rule, namely $0000\cdots\cdots$.

Next consider the problem of periodic solutions, i.e. states which satisfy

$$b = N^k(b)$$

for some $k \geq 1$, of course, if $k = 1$ we simply get fixed points, i.e. orbits of period $1$. Consider first the case of period $2$ orbits, so that we require fixed points of the map $N^2$. We can solve this problem as in the case of $k = 1$. To illustrate the method, suppose again that $p = 3$, i.e. 3-bit rule $R_3$. Then $N^2$ is defined by a 5-bit rule $R_5$ defined by

$$R_5(b_0b_1b_2b_3b_4) = R_3(R_3(b_0b_1b_2), R_3(b_1b_2b_3), R_3(b_2b_3b_4)) \tag{2.2}$$

Again we find all the 5-bit binary strings for which

$$R_5(b_0b_1b_2b_3b_4) = b_2$$

and form a graph with these as vertices $v_1,\cdots,v_k$ and we add an edge from $v_i$ to $v_j$ if $v_i$, $v_j$ are of the form

$$
\begin{aligned}
v_i: &\quad b_0^i b_1^i b_2^i b_3^i b_4^i \\
v_j: &\quad \quad b_0^j b_1^j b_2^j b_3^j b_4^j
\end{aligned}
\tag{2.3}
$$

where $b_l^j = b_{l+1}^i$, $0 \le l \le 3$.

Fixed states can now be read from the graph by writing down infinite sequences of linked vertices, if they exist. Vertices satisfying (2.3) will be called *linkable*. Because of the boundary conditions the sequences of linked vertices must begin at *initial vertices*. These are vertices which satisfy

$$R_5\left(0\ 0\ b_2 b_3 b_4\right) = R_3\left(0\ , R_3\left(0\ b_2 b_3\right), R_3\left(b_2 b_3 b_4\right)\right)$$

instead of (2.2). (We must set the first term on the right hand side to zero, rather than $R_3\left(0\ 0\ b_2\right)$, because of the boundary condition.) Thus the graph which includes the initial vertices may have more vertices than the one generated just by the rule (2.2). (Of course, similar remarks hold for higher order periodic solutions.) In general, we have

**Theorem 2.1** A dynamical system defined by a cellular rule $R_p$ of length $p$ (where $p$ is odd) has a periodic solution of period $k$ if there is a path of infinite length (i.e. at least one loop) in the graph consisting of vertices $v_i$ corresponding to binary strings of length $k(p-1)+1$ for which the rule $R_{k(p-1)+1}$ given inductively by

$$
\begin{aligned}
R_{k(p-1)+1}\left(b_0 \cdots b_{k(p-1)}\right) = R_p\Big(&R_{(k-1)(p-1)+1}\left(b_0 \cdots b_{(k-1)(p-1)}\right), \\
&R_{(k-1)(p-1)+1}\left(b_1 \cdots b_{(k-1)(p-1)+1}\right), \cdots, R_{(k-1)(p-1)+1}\left(b_{p-1} \cdots b_{k-1}\right)\Big)
\end{aligned}
$$

satisfies

$$R_{k(p-1)+1}\left(b_0 \cdots b_{k(p-1)}\right) = b_{k(p-1)/2}\,.$$

Edges of the graph join linkable vertices. □

**Example 2.2**consider the 3-bit rule

| | $R_3$ |
|---|---|
| 000 | 1 |
| 001 | 0 |
| 010 | 0 |
| 011 | 0 |
| 100 | 0 |
| 101 | 0 |
| 110 | 1 |
| 111 | 0 |

(2.4)

To find period-2 orbits, we must determine the corresponding 5-bit rule: this is given by

| | $R_5$ |
|---|---|
| 00000 | 0 |
| 00001 | 1 |
| 00010 | 0 |
| 00011 | 0 |
| 00100 | 1 |
| 00101 | 1 |
| 00110 | 0 |
| 00111 | 1 |
| 01000 | 0 |
| 01001 | 1 |
| 01010 | 1 |
| 01011 | 1 |
| 01100 | 0 |
| 01101 | 0 |
| 01110 | 0 |
| 01111 | 1 |
| 10000 | 0 |
| 10001 | 0 |
| 10010 | 1 |
| 10011 | 1 |
| 10100 | 1 |
| 10101 | 1 |
| 10110 | 0 |
| 10111 | 1 |
| 11000 | 0 |
| 11001 | 0 |
| 11010 | 0 |
| 11011 | 0 |
| 11100 | 0 |
| 11101 | 0 |
| 11110 | 0 |
| 11111 | 1 |

The initial vertices are given by:

| | $R_5$ |
|---|---|
| 00000 | 0 |
| 00001 | 0 |
| 00010 | 1 |
| 00011 | 1 |
| 00100 | 1 |
| 00101 | 1 |
| 00110 | 0 |
| 00111 | 1 |

The rule values which fix the middle term are shown in the graph below, together with directed edges between linkable vertices.

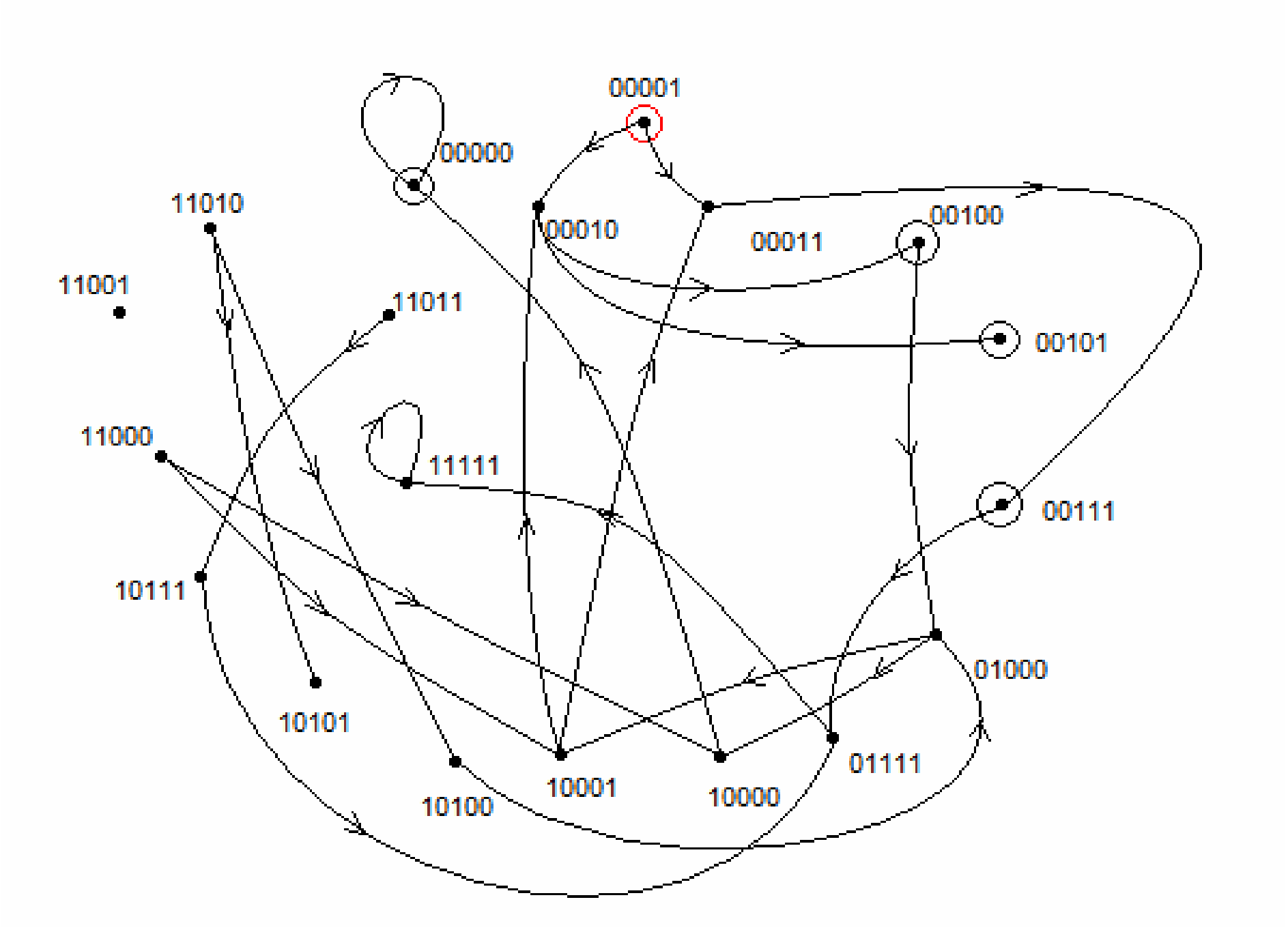

From the graph we see that the period-2 orbits are of the form

00.1111111111111111111······

00.0000000000000000000······

00.0011111111111111111······

00.0010000000000000000······

00.0010001000100010001······

00.0010001000000000000······

00.0010001000100000000······

·····································

00.0010001111111111111······

00.0010001000111111111······

·····································

00.100010001000100010001······

00.100000000000000000000······

00.100010000000000000000······

00.100010001000000000000······

·····································

00.100011111111111111111······

00.100010001111111111111······

……………………………………
(assuming zero boundary conditions).
In the following part we show that periodic orbits can be found easily by interpreting the binary states as real numbers in the interval $[0,1)$.

Consider again the case of semi-infinite strings with zero left-hand boundary condition, i.e. strings of the form

$$\underbrace{0\cdots0\ 0}_{p-1/2}\vdots\ b_1\ b_2\ b_3\cdots\cdots$$

As before, let $B$ denote the set of all such strings. We shall define a map $v$ from $B$ to the semi-open unit interval $[0,1)\subseteq R$ by

$$v\left(b_1\ b_2\ b_3\ \cdots\right)=\sum_{i=1}^{\infty}b_i 2^{-i}\ ,$$

i.e. by equating a string with the corresponding 2-adic representation of a real number, where $b_i$ is the binary bit in the ith place. This map is not, of course, one-to-one. Certain strings map to the same number; in fact, any string ending in an unbroken infinite string of 1's represents the same number in $[0,1)$ as another one ending in an infinite number of zeros. Thus, for example, the stings

$$0101101011111111111111111\cdots\cdots$$
$$0101101100000000000000000\cdots\cdots$$

represent the same number, i.e. $2^{-2}+2^{-4}+2^{-5}+2^{-7}+2^{-8}=91/256$. The exception is the string of all 1's, i.e. 11111111⋯⋯, which we must exclude from $B$ since it represents the number 1 and we have insisted on the left-hand boundary condition 0. Clearly, for all other strings (i.e. those not ending in unbroken infinite strings of 0's or 1's) the map $v$ is one-to-one.

**Lemma 2.2** The set $B$ (minus the infinite string 1111111⋯) is in 1-1 correspondence with the Cantor Set $C$ (minus the number 1). Moreover, the set of numbers $\alpha\in[0,1)$ for which the set $v^{-1}(\alpha)$ has two elements is countable.

**Proof:** For the first part note that the map $c:B\,|\,(1111\cdots)\to C\setminus\{1\}$ given by

$$c\left(b_1\ b_2\ b_3\cdots\right)=\sum_{i=1}^{\infty}b_i'3^{-i}$$

where

$$b_i' = \begin{cases} 0 & if\ \ b_i = 0 \\ 2 & if\ \ b_i = 1 \end{cases}$$

is a 1-1 map. For the second statement consider for each point $\alpha \in [0,1)$ such that $\#\left(v^{-1}(\alpha)\right) = 2$ the representation ending in an unbroken string of 0's. Such numbers are clearly rational and so the result follows. □

Note that, of course, not all rational number can be expressed as binary strings ending in 0's; for example, 1/3 is the periodic string $01010101\cdots\cdots$. We will denote by $B_0$ the subset of $B$ obtained by removing all double points of the map $v$, i.e. all binary strings ending with all 0's or all 1's. Hence the map

$$v_0 : \ B_0 \to [0,1)$$

given by $v_0 = v \mid B_0$ is 1-1. We will put on $B_0$ the topology induced from the usual topology of $[0,1)$ by $v_0^{-1}$. Now let $N$ be a nonlinear map from $B_0$ to $B$ determined by a p-bit rule. (We shall denote the corresponding map from $B$ to $B$ by the same letter $N$. No confusion should arise here.) Then the map $v_0$ induces a map $\tilde{N} : \ [0,1) \setminus \Gamma \to [0,1)$, where $\Gamma$ is the set of all numbers with a binary representation ending in 0's, which make the diagram

$$\begin{array}{ccc} B_o & \xrightarrow{v_o} & [0,1) \setminus \Gamma \\ N \downarrow & & \downarrow \tilde{N} \\ B & \xrightarrow{v} & [0,1) \end{array}$$

commutative.

If we allow two-valued maps, we can extend this diagram to the following one:

$$\begin{array}{ccc} B & \xrightarrow{v} & [0,1) \\ N \downarrow & & \downarrow \tilde{N} \\ B & \xrightarrow{v} & [0,1) \end{array} \tag{2.5}$$

But this time, $\tilde{N}$ may be two valued on $\Gamma$.

**Lemma 2.3** There exists a countable subset $S \subseteq B$ such that the diagram

$$\begin{array}{ccc} B \setminus S & \xrightarrow{v} & [0,1) \\ N \ \downarrow & & \downarrow \tilde{N} \mid_{v(B \setminus S)} \\ B \setminus S & \xrightarrow{v} & [0,1) \end{array}$$

is well-defined, i.e. the induced map $\tilde{N} \mid_{v(B \setminus S)}$ is 1-1.

**Proof** Let $S_0 = B \setminus B_0$. Then $\tilde{N}$ may be double valued on $S_0$. Let $\tilde{N}(T) = \bigcup_{t \in T} \tilde{N}(t)$, for any set $T \subseteq B$. Then we define inductively the sets

$$S_i = \tilde{N}(S_{i-1}), \quad i \geq 1.$$

and

$$S = \bigcup_{i \geq 1} S_i.$$

Clearly $S$ satisfies the statement of the lemma (since $S$ is clearly countable). □

We may assume therefore that, modulo a countable set, the diagram in (2.5) is well-defined. As a function from $[0,1)$ to itself, we can draw a graph of the function $\tilde{N}$. For example, for the rule (2.1) we obtain the graph shown in fig.2. Note that fixed points of the map are given by fixed points of $\tilde{N}$, i.e. real number $x$ such that $x = \tilde{N}(x)$ —these are just points on the diagonal.

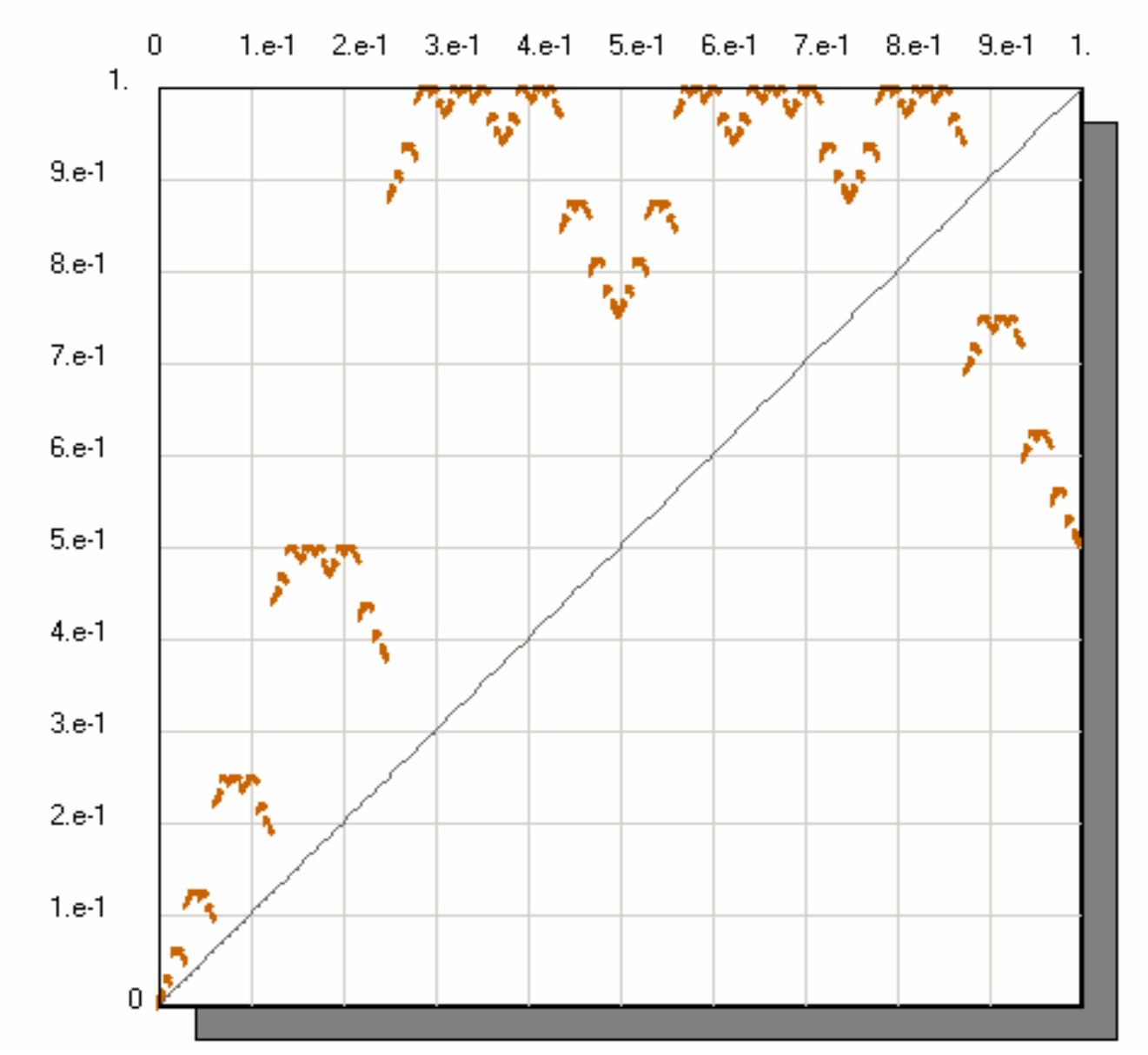

fig2.

We can also determine periodic points in the original cellular system by examining the iterated maps $\tilde{N}^k$ for each $K \geq 1$. Points on the diagonal of the graph of $\tilde{N}^k$ clearly represent periodic points of period $k$ in the cellular automaton. For the rule (2.4), the graph of $\tilde{N}^2$ is shown in fig.3, which shows that the period 2 points are as shown in Example 2.2. Higher order periodic solutions are easily found by evaluating $\tilde{N}^k$ for some $k$. $\tilde{N}^5$, for example, if shown in fig.4.

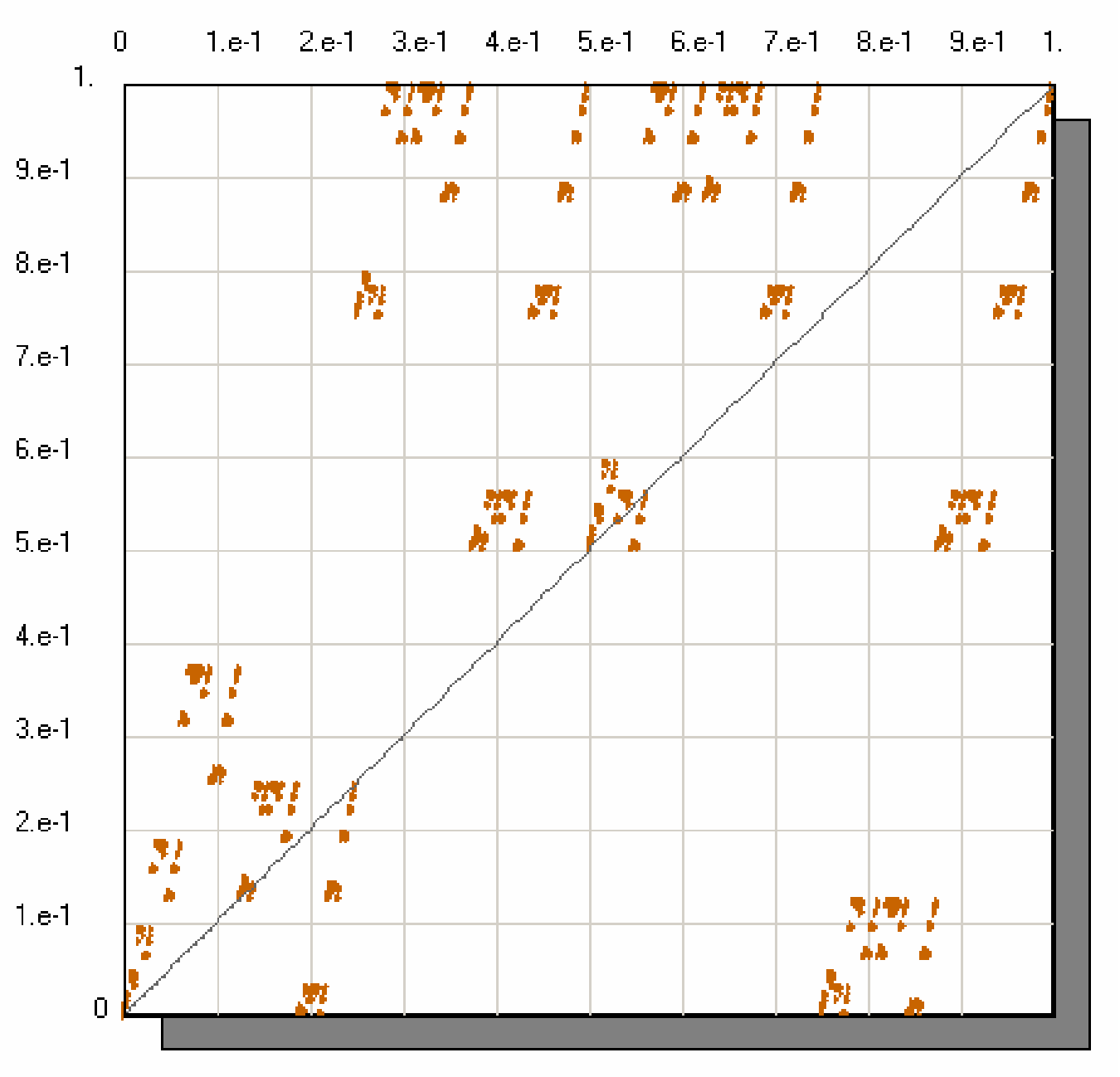

fig.3.

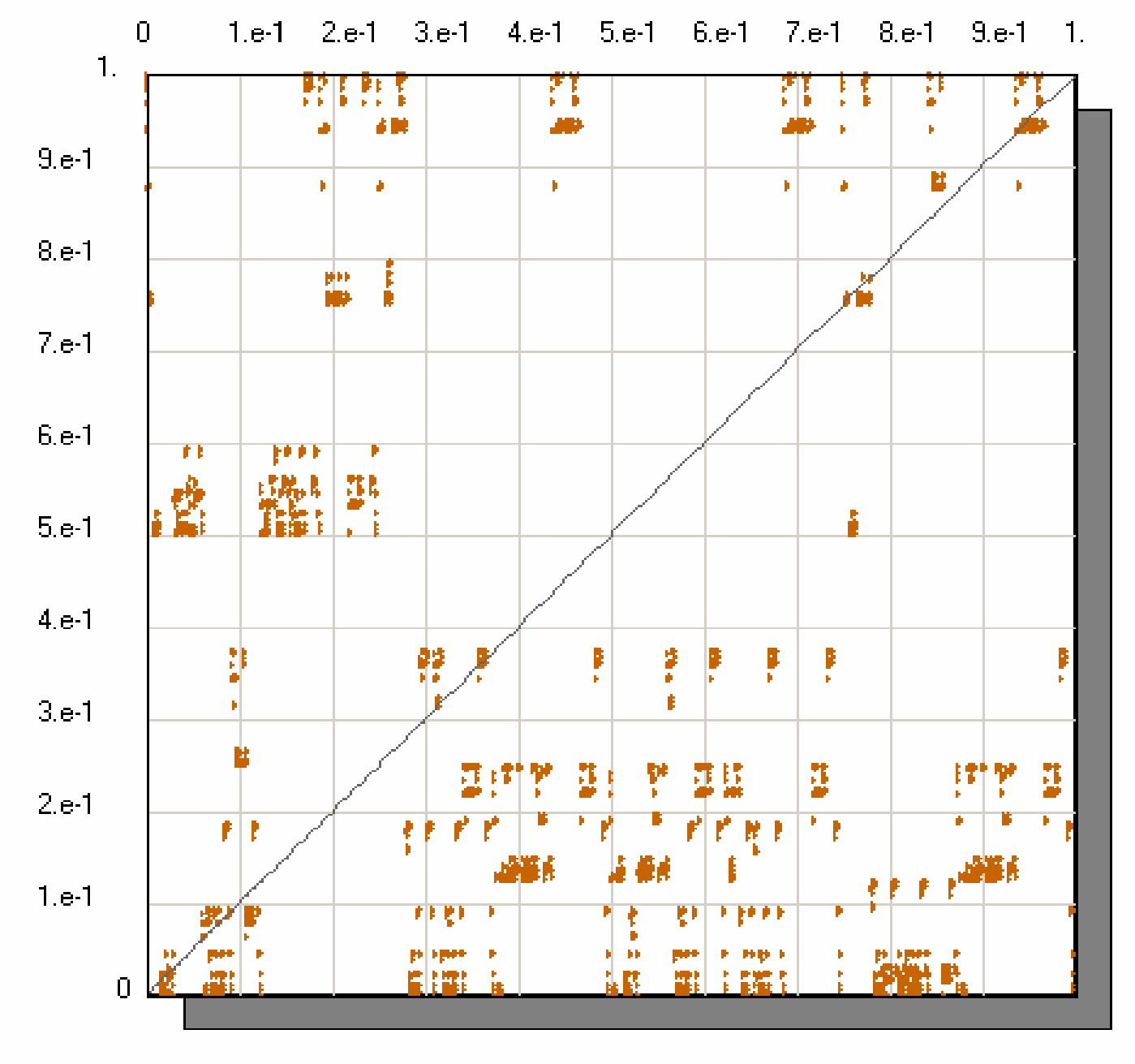

fig.4.

Consider next the continuity properties of the map $\tilde{N}$. We have the following result.

**Lemma 2.4** $\tilde{N}:[0,1)\setminus v(S)\to[0,1)\setminus v(S)$ is continuous.

**Proof** Any point in $[0,1)\setminus v(S)$ has a binary (2-adic) representation which does not end in an unbroken string of 0's or 1's. Let $b\in[0,1)\setminus v(S)$, so that $b=(b_i)_{i\geq 1}$, where there exist 0's and 1's in the $b_i$ for arbitrarily large $i$. Let $\varepsilon>0$ be 'small' and let $\varepsilon_1\varepsilon_2\varepsilon_3\cdots$ be its 2-adic representation. Then for some $K$, $\varepsilon_i = 0$, $1 \le i \le K$. For any $K_1$ there exists $J \ge K_1$ such that $b_J = 0$, i.e. $b$ is of the form

$$\begin{array}{c} \cdots b_{J-1}\ b_J\ b_{J+1} \\ \cdots\cdots \quad 0\cdots\cdots \\ J \end{array}$$

If we choose $K > K_1$, then $b+\varepsilon$ will only be different from $b$ after the $(J-1)'th$ place. Hence, for a p-bit rule, $\tilde{N}(b+\varepsilon)$ will only differ from $\tilde{N}(b)$ after the $(J-(p-1)/2)$ place. Since we may take $K_1$ arbitrarily large we see that $\lim_{\varepsilon \to 0^+} \tilde{N}(b+\varepsilon) = \tilde{N}(b)$. A similar argument applies for $b-\varepsilon$ and the result follows. □

We can also ask: when is $N$ differentiable?

$\tilde{N}$ is differentiable when

$$\lim_{\varepsilon \to 0} \frac{\tilde{N}(b+\varepsilon) - \tilde{N}(b)}{\varepsilon}$$

exists, of course. Suppose that $b$ is run limited, i.e. that in the binary representation of $b$ any consecutive sequences of 1's or 0's is bounded. If this bound is $K$, then no sequence of 1's (or 0's) in $b$ can be greater in length than $K$. Thus, for example,

$$b = \ 001110011100111001111\cdots\cdots$$

has run length 3. Suppose $\varepsilon > 0$ and 'small', if $b$ is run length limited by $K$ then it is of the form

$$b = \ \cdots\cdots 11\cdots 100\cdots 011\cdots 100\cdots 011\cdots$$

where each block of 1's or 0's is of length $\le K$.

For simplicity, consider a 3-bit rule, and suppose $b+\varepsilon$ is of the form so that $b+\varepsilon$ and $b$ have the same binary representations up to the $Lth$ place. For a 3-bit rule, $\tilde{N}(b+\varepsilon)$ and $\tilde{N}(b)$ may therefore differ only after the $(L-1)th$ place, so that

$$\left|\frac{\tilde{N}(b+\varepsilon)-\tilde{N}(b)}{\varepsilon}\right| \le 2^{K+1}.$$

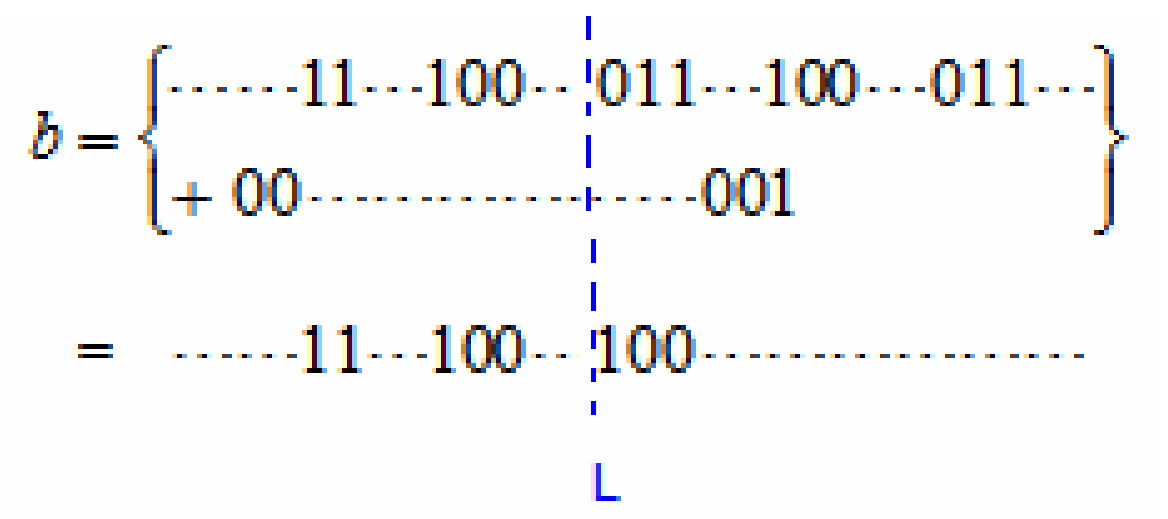

Similar remarks apply if $\varepsilon < 0$ ( here we consider the 'worst case' of subtracting 1 from a string of 0's ). For a p-bit rule (p odd), we therefore have

$$\left|\frac{\tilde{N}(b+\varepsilon)-\tilde{N}(b)}{\varepsilon}\right| \le 2^{K+(p-1)/2}.$$

Hence we have proved

**Theorem 2.2** The map $\tilde{N}:\ [0,1)\to[0,1)$ associated with a p-bit cellular automaton is differentiable at the points which have run length limited binary representations and the derivative is bounded by $2^{K+(p-1)/2}$, where $K$ is the run length of the binary representation of the point. □

**Remark** The number of run-length limited points is clearly countable, whereas, as seen above, the set of points of continuity is uncountable. □

## 3. Two-dimensional Automata

Two dimensional automata are nonlinear dynamical systems defined on two-dimensional arrays of binary bits. We shall assume that the state-space consists of doubly-infinite matrices of binary bits $b=\left(b(i,j)\right)_{1\le i<\infty,\ 1\le j<\infty}$. The dynamics is again defined by local rules, which we shall take as being defined on $3\times 3$ neighbourhoods of each bit in the state. Thus a <u>rule</u> is a map $R:\ M_{3\times 3}\to\{0,1\}$ where $M_{3\times 3}$ is the set of $3\times 3$ binary matrices, and the dynamics $N$ are defined by

$$b' = N(b)$$

where

$$b'(i,j) = R\left(b_M\,(i,j)\right)$$

and $b_{M(i,j)}$ is the $3\times 3$ submatrix of $b$ with central element $b(i,j)$. We shall take zero boundary conditions, so that

$$b(i,j)=0 \quad \text{if} \quad i=0 \ \text{ or } \ j=0.$$

A typical example is the well-known game of life of Conway. This is defined by the rule

$$R\left(\beta=\begin{pmatrix}\beta_1 & \beta_2 & \beta_3\\ \beta_4 & \beta_5 & \beta_6\\ \beta_7 & \beta_8 & \beta_9\end{pmatrix}\right)=\begin{cases}1 & \begin{aligned}&if\ \left(\beta_5=1\ and\ \ \#\beta=2\ or\ 3\right)\\ &or\ \left(\beta_5=0\ and\ \ \#\beta=3\right)\end{aligned}\\ 0 & otherwise\end{cases}$$

where $\#\beta = \sum_{i=1}^{9} \beta_i$ .

As in the one-dimensional case, we shall first consider the problem of determining fixed points. Thus, again as before, we consider all $3 \times 3$ matrices $\beta$ of binary bits such that

$$R(\beta) = \beta_5$$

i.e. matrices in the domain of the rule which fix the middle bit. Again we enumerate the matrices $\beta$ with this property and denote them by $v_1, \cdots, v_k$ . We shall 'fit the $3 \times 3$ matrices together' to form a complete state first horizontally and then vertically. Hence, as in the one-dimensional case, we form a graph with vertices $v_1, \cdots, v_k$ and edges $v_i \to v_j$ if $v_j$ can be placed 'on the right' of $v_i$ . By this we mean that if

$$v_i = \begin{pmatrix} \beta_1 & \beta_2 & \beta_3 \\ \beta_4 & \beta_5 & \beta_6 \\ \beta_7 & \beta_8 & \beta_9 \end{pmatrix}, \qquad v_j = \begin{pmatrix} \gamma_1 & \gamma_2 & \gamma_3 \\ \gamma_4 & \gamma_5 & \gamma_6 \\ \gamma_7 & \gamma_8 & \gamma_9 \end{pmatrix}$$

then

$$\begin{pmatrix} \beta_2 & \beta_3 \\ \beta_5 & \beta_6 \\ \beta_8 & \beta_9 \end{pmatrix} = \begin{pmatrix} \gamma_1 & \gamma_2 \\ \gamma_4 & \gamma_5 \\ \gamma_7 & \gamma_8 \end{pmatrix}.$$

Next we find all infinite paths in the graph which begin with a vertex of the form

$$v = \begin{pmatrix} 0 & \square & \square \\ 0 & \square & \square \\ 0 & \square & \square \end{pmatrix},$$

(for zero boundary conditions). (Vertices of this form will be labeled as initial vertices.) The first horizontal band of the state must have vertices of the form

$$v = \begin{pmatrix} 0 & 0 & 0 \\ \square & \square & \square \\ \square & \square & \square \end{pmatrix}$$

(or $\begin{pmatrix} 0 & 0 & 0 \\ 0 & \square & \square \\ 0 & \square & \square \end{pmatrix}$ for the first one). The set of infinite paths which have just been found form the vertices $W_1, \cdots W_L$ of a second graph. An edge $W_i \to W_j$ exists if $W_j$ can be placed 'below' $W_i$ in an obvious sense (starting from a correct initial point). We then search for infinite paths in the second graph, and these correspond to semi-infinite fixed points.

**Example 3.1** Consider the 5-bit rule:

| $b_1$ $b_2$ $b_3$ $b_4$ $b_5$ $\to$ $b_1\, b_2\, b_3\, b_4\, b_5$ | $R_5$ |
|---|---|
| 00000 | **1** |
| 00001 | **1** |
| 00010 | **1** |
| 00011 | **0** |
| 00100 | **1** |
| 00101 | **0** |
| 00110 | **0** |
| 00111 | **0** |
| 01000 | **1** |
| 01001 | **1** |
| 01010 | **1** |
| 01011 | **0** |
| 01100 | **0** |
| 01101 | **0** |
| 01110 | **0** |
| 01111 | **0** |
| 10000 | **1** |
| 10001 | **1** |
| 10010 | **1** |
| 10011 | **0** |
| 10100 | **0** |
| 10101 | **0** |
| 10110 | **0** |
| 10111 | **0** |
| 11000 | **1** |
| 11001 | **1** |
| 11010 | **1** |
| 11011 | **0** |
| 11100 | **0** |
| 11101 | **0** |
| 11110 | **0** |
| 11111 | **0** |

We could use tow graphs which we have illustrated before to find the fixed points according to this 5-bit rule:

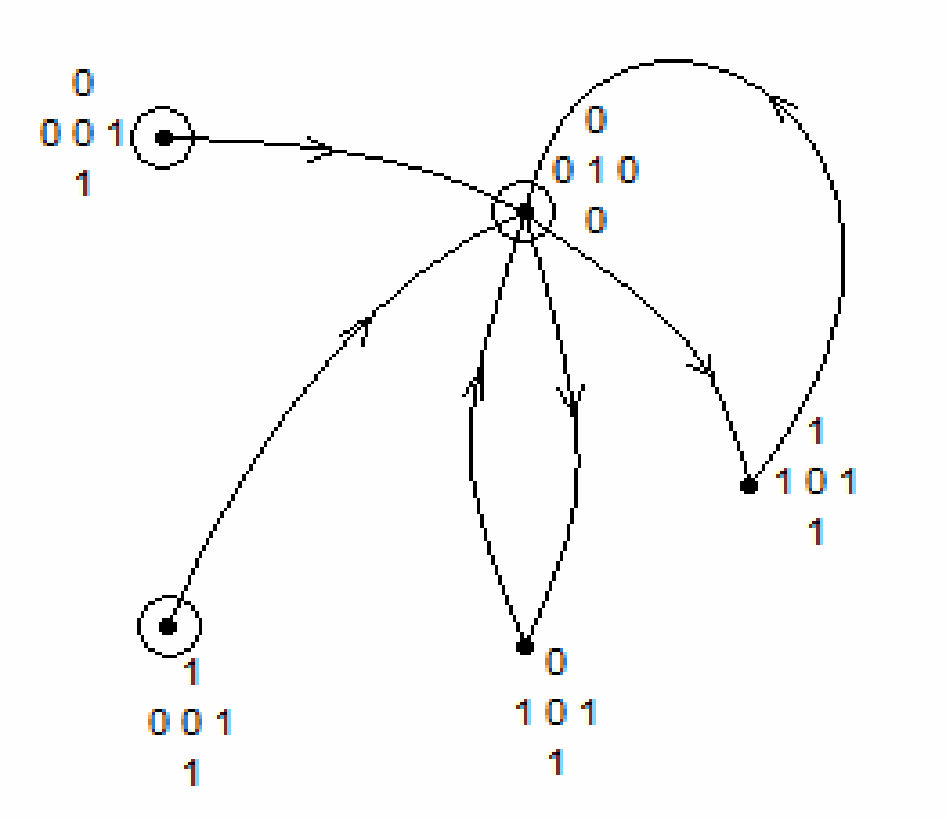

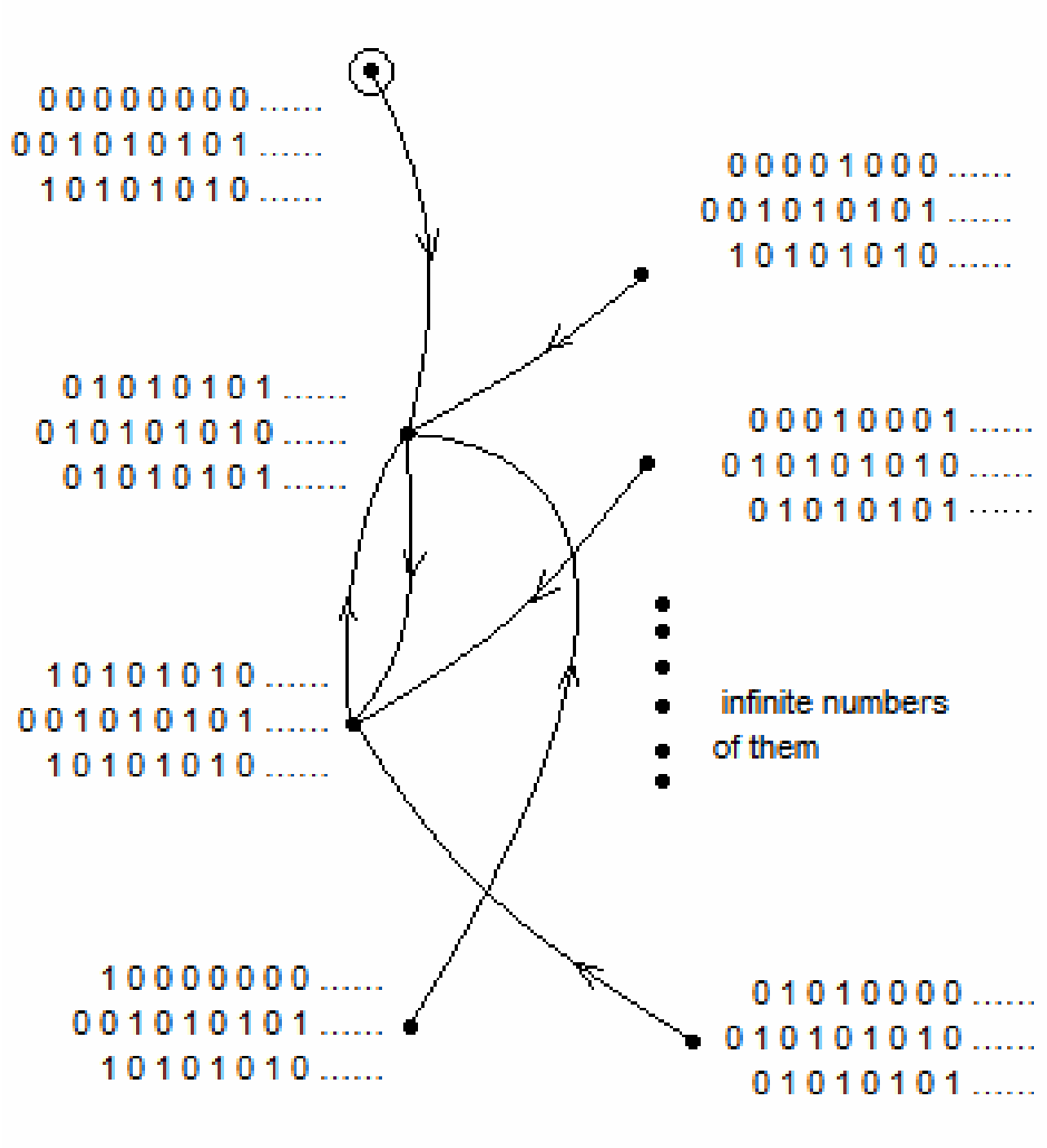

Clearly, in this example there is only one fixed point:

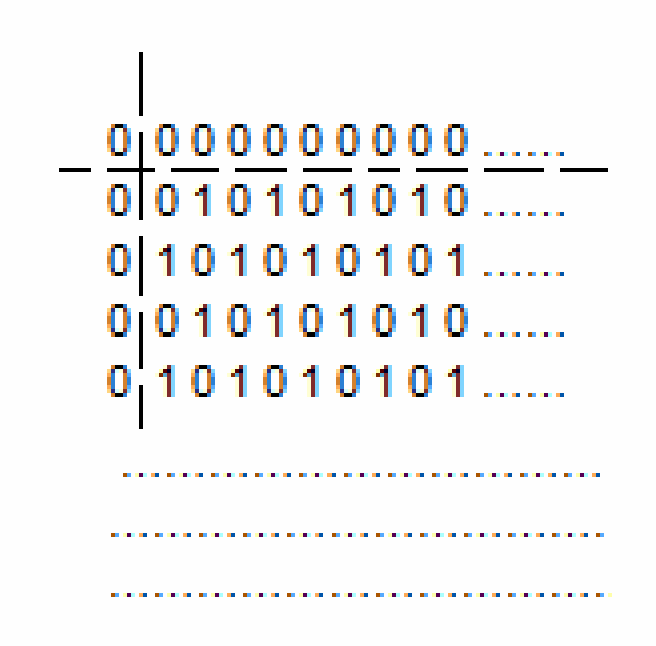

The above method for finding fixed points can be generalized to find periodic solutions as in the one-dimensional case, but it is clear computationally difficult. In the last part we will show how to generalize the 2-adic method for one-dimensional automata to the two-dimensional case. To do this we first map the two-dimensional state matrix onto a binary string. Therefore consider a semi-infinite two-dimensional binary matrix $A$, corresponding to a cellular state.
To $A$ we associate the binary string

$$b = b_1 b_2 b_3 \cdots$$

where $b_1 = A(1,1)$

and $$b_{n(n-1)+l} = \begin{cases} A(l,\ n),\ 1 \le l \le n \\ A(n,\ l-n),\ n+1 \le l \le 2n-1 \end{cases}.$$

(For example, see fig.5.)

$$\begin{array}{cccccccccc} b_1 & b_2 & b_5 & b_{10} & \cdots & \cdots & \cdots & \cdots & \cdots & \cdots \\ b_4 & b_3 & b_6 & b_{11} & \cdots & \cdots & & & & \\ b_8 & b_9 & b_7 & b_{12} & \cdots & \cdots & A & & & \\ b_{14} & b_{15} & b_{16} & b_{13} & \cdots & \cdots & & & & \\ \vdots & \vdots & \vdots & \vdots & \ddots & \cdots & & & & \\ \vdots & \vdots & \vdots & \vdots & \vdots & \ddots & & & & \end{array}$$

fig.5

**Remark** We could choose any other way to count the binary bits in $A$.

Now we associate with the binary string $b$, the real number $\nu(b_1 b_2 b_3 \cdots) = \sum_{i=1}^{\infty} b_i 2^{-i}$ (as in the one-dimensional case). The remaining theory follows as in the one-dimensional case (i.e. we can define the functions $N$ and $\tilde{N}$ in the same way). Hence we will simply give an example of periodic solutions in the game of life (see fig.6. and fig.7. ). States of period 5 appear on the diagonal of the map $\tilde{N}^5$ - a period 5 orbit is shown in fig.7.

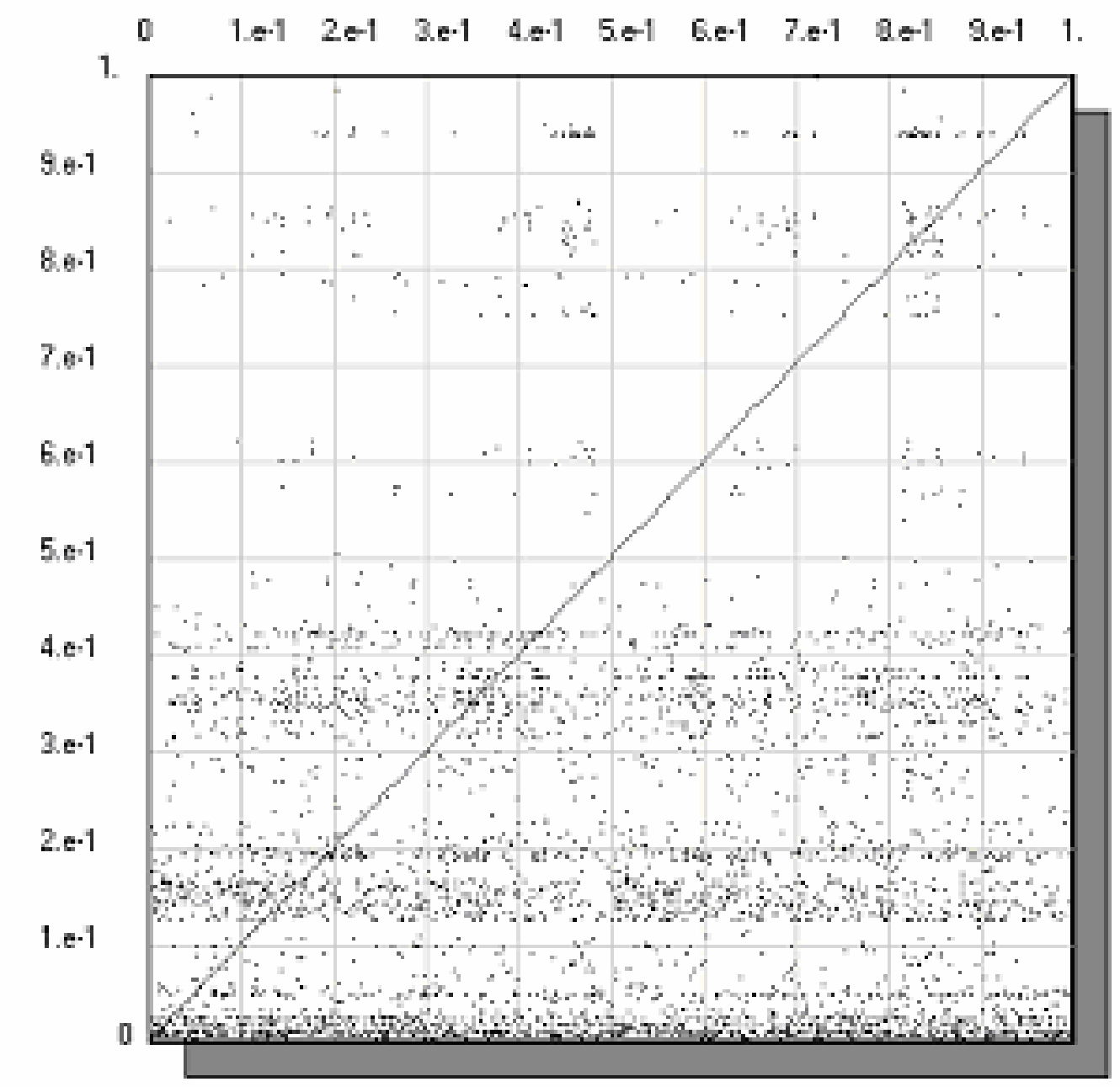

fig.6. Game of Life Function ( $\tilde{N}^5$ )

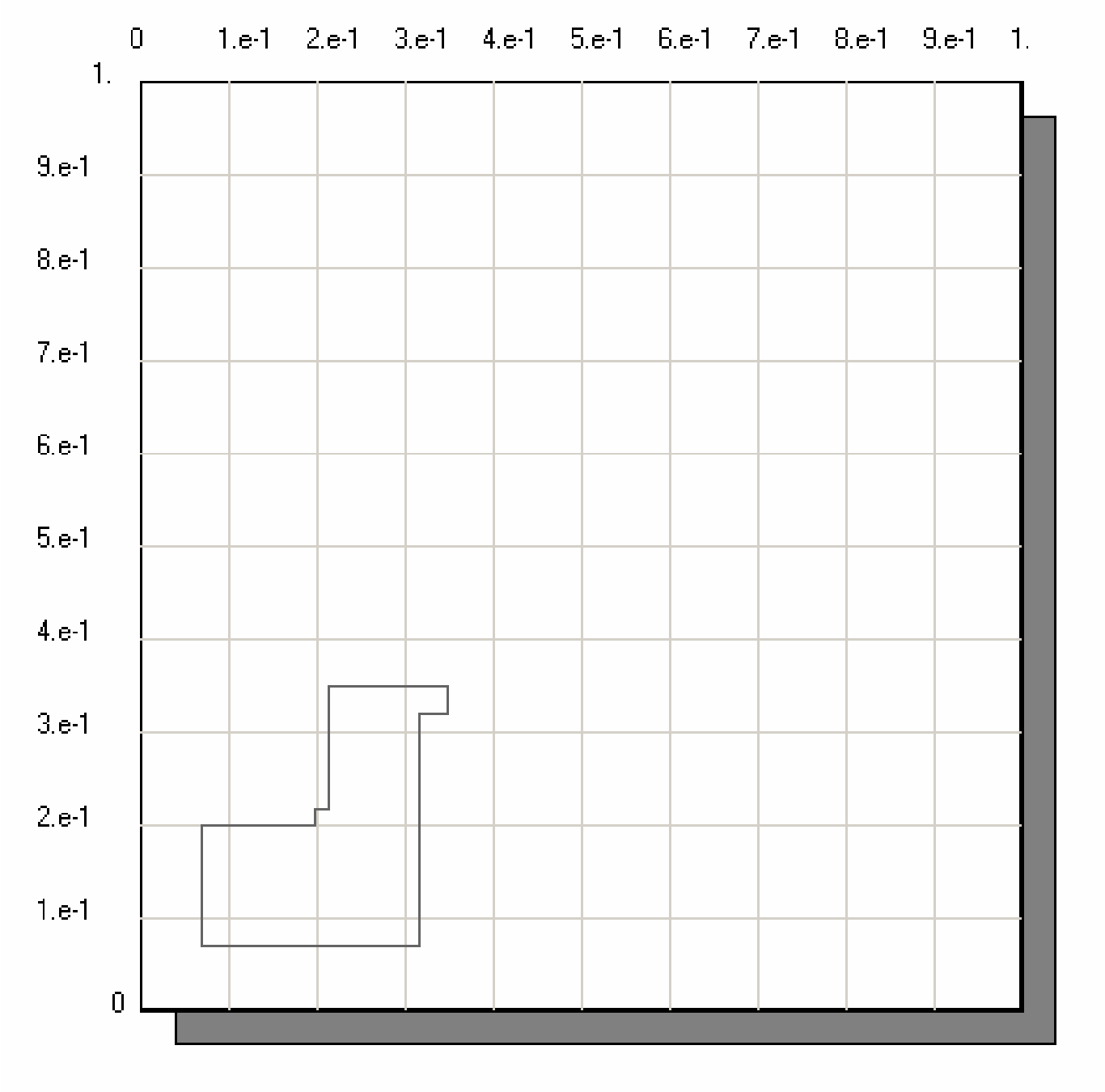

fig.7. Periodic orbit of period 5

## 4. Entropy

Entropy is a measure of the amount of variability in the dynamics, and shows whether the dynamics are regular or chaotic. The entropy in symbolic dynamical systems can be evaluated easily by using the Perron-Frobenius theory (see, eg. [Lind and Marcus]). In the case of cellular automata, it is somewhat more difficult to compute. We shall show how to use the map $\tilde{N}$ to determine the entropy in such systems. First we define the notion of entropy in a general dynamical system. Thus, let $M$ be a compact manifold with metric $\phi$ and $\phi$ a dynamical system defined on $M$ . A subset $E \subseteq M$ is $(n,\ \varepsilon)$- spanning for $\phi$ if, for all $x \in M$ , there exists $y \in E$ such that

$$\rho\left(\phi^j(x),\ \phi^j(y)\right) \le \varepsilon, \qquad \text{for all } 0 \le j \le n .$$

By continuity of $\phi$ and compactness of $M$ , there exists a finite $(n,\ \varepsilon)$- spanning set for all $\varepsilon > 0$ and $n \ge 1$. Let $k_n(\phi,\ \varepsilon)$ be the number of elements in a minimal $(n,\ \varepsilon)$- spanning set. Then we define the *entropy* of $(M,\ \phi)$ by

$$h(\phi) = \lim_{\varepsilon \to 0} \limsup_{n \to \infty} \frac{1}{n} \log k_n(\phi,\ \varepsilon).$$

Given a cellular automaton $N$ we associate with it the map $\tilde{N} : [0,1) \to [0,1)$ as above

(restricted to the set of numbers with unique binary representation) and we consider, for each $n \geq 1$, the map

$$\tilde{N}^{(n)}: \left[0,1\right) \rightarrow \underset{i=1}{\overset{n}{X}} \left[0,1\right)$$

given by

$$\tilde{N}^{(n)}\left(x\right) = \left(x, \tilde{N}\left(x\right), \cdots, \tilde{N}^{n-1}\left(x\right)\right).$$

Give $\underset{i=1}{\overset{n}{X}} \left[0,1\right)$ the metric defined by

$$\rho^{(n)}\left(\left(\xi_1, \cdots, \xi_n\right), \left(\eta_1, \cdots, \eta_n\right)\right) = \max_{i=1,\cdots,n} \left|\xi_i - \eta_i\right|.$$

Then the set $E \subseteq \left[0,1\right)$ is $\left(n,\ \varepsilon\right)$- spanning for $\tilde{N}$ if and only if it is $\varepsilon$ - spanning for $\tilde{N}^{(n)}$ in the sense that, for any $x \in \left[0,1\right)$ there exists $y \in E$ such that

$$\rho^{(n)}\left(\tilde{N}^{(n)}\left(x\right), \tilde{N}^{(n)}\left(y\right)\right) \leq \varepsilon.$$

To compute approximations to the entropy for a cellular automaton we can therefore use the following procedure:

(i). Set $n = 1,\ \varepsilon = 0.1$ (say), choose a finite number of 'well-distributed' points in $\left[0,1\right)$ and number them $x_1, \cdots, x_k$.

(ii). Find a minimal $\varepsilon$ - spanning set $E$ by choosing $x_1$ first and then testing inductively to see if $x_i$ belongs to $E$ given that $x_{i_1}, \cdots, x_{i_l} \in E$. Thus if

$$\rho^{(n)}\left(\tilde{N}^{(n)}\left(x_{ij}\right), \tilde{N}^{(n)}\left(x_i\right)\right) \leq \varepsilon$$

for some $j \in \left\{1, \cdots, l\right\}$, then $x_i \notin E$. Otherwise $x_i \in E$.

(iii). Compute

$$h_{(n,\varepsilon)} = \frac{1}{n} \log\#(E).$$

(iv). Let $n+1 \rightarrow n$ and $\varepsilon = \varepsilon/2$ and return to (i), until $h_{(n,\varepsilon)}$ converges.

The entropy is easy to compute numerically from the above considerations.

## 5. Conclusions

In this paper we have seen how to study some of the interesting dynamics of complex cellular systems - in particular, we have developed efficient methods for finding fixed points and periodic orbits in one- and two- dimensional systems. Clearly, these methods generalise to higher-dimensional systems, but the 2-adic representations then require high precision arithmetic in the real number equivalents. (This is, of course, no problem for modern computer algebra packages, although the computation time will be larger for higher-dimensional systems.) The 'complexity' of cellular systems can be defined in terms of the entropy of the associated function $\tilde{N}$ and a simple computational algorithm has been given.

In a future paper we shall study the stability of periodic solutions and the existence of connecting orbits.